\documentclass{article}

\usepackage{arxiv}

\usepackage[utf8]{inputenc}
\usepackage[T1]{fontenc}
\usepackage{hyperref}
\usepackage{url}
\usepackage{booktabs}
\usepackage{amsfonts}
\usepackage{amsmath}
\usepackage{amssymb}
\usepackage{nicefrac}
\usepackage{microtype}
\usepackage{graphicx}
\usepackage{natbib}
\usepackage{doi}
\usepackage{xcolor}
\usepackage{multirow}
\usepackage{array}
\usepackage{enumitem}
\usepackage{caption}
\usepackage{subcaption}

\title{Probe-and-Refine Tuning of Repository Guidance\\for Coding Agents}

\author{
Asa Shepard \\
Williams College \\
\texttt{as66@williams.edu}
\And
Jeannie Albrecht \\
Williams College \\
\texttt{jra1@williams.edu}
}

\renewcommand{\shorttitle}{Probe-and-Refine Tuning of Repository Guidance}

\hypersetup{
pdftitle={Probe-and-Refine Tuning of Repository Guidance for Coding Agents},
pdfauthor={Asa Shepard, Jeannie Albrecht},
pdfkeywords={coding agents, SWE-bench, AGENTS.md, repository context, probe-and-refine, inference-time adaptation},
}

\begin{document}
\maketitle

\begin{abstract}
LLM-based coding agents need higher-level operational knowledge about a repository (which files house which subsystems, how to run the test suite, which workflows have historically led to wrong fixes) that does not exist in the code itself.
Engineers typically maintain \texttt{AGENTS.md} files to supply this context as instructions for coding agents, but whether they help is contested: recent studies disagree on whether LLM-generated guidance improves or harms agent performance.
In this paper we show that how the guidance is produced is the decisive variable, and introduce \emph{probe-and-refine tuning}: a procedure that uses synthetic bug-fix probes to iteratively diagnose and patch a repository's guidance file through single-shot LLM calls, with no agent loop or tool use during tuning.
On SWE-bench Verified across four independent trials with Qwen3.5-35B-A3B at 200 steps, probe-and-refine achieves 33.0\,\% mean resolve rate vs.\ 28.3\,\% for the static knowledge base used to initialize it and 25.5\,\% for an unguided baseline ($p < 0.001$ for both probe-and-refine contrasts).
The improvement comes from coverage rather than precision: refined guidance produces evaluable patches for 14.5 percentage points (pp) 
more instances while per-patch precision remains statistically constant ($\sim$59\,\%, $p = 0.119$), showing that improved guidance helps agents reach the correct file rather than improving the quality of the changes they make.
Further, a step-budget experiment shows that guidance is what lets the agent use a larger step budget productively, and a 
cross-model experiment with NVIDIA-Nemotron-3-Nano-30B-A3B finds that the tuning loop degrades when the model cannot generate sufficiently diagnostic output, though per-patch precision remains constant even then.
\end{abstract}

\section{Introduction}
\label{sec:intro}

LLM-based coding agents operate in a loop of reading code, running commands, and proposing edits to code.  They are increasingly being deployed and used for software engineering tasks in real-world contexts.
The effectiveness of the agents depends on more than just the model's underlying capability.
A code repository is a body of accumulated decisions: subsystem boundaries, naming conventions, debugging workflows that experienced contributors have internalized but that are not documented in the code itself.
A model that has never worked in a particular codebase has to reconstruct this implicit knowledge on the fly, often by trial and error, and frequently lands in the wrong file or proposes a fix that breaks orthogonal subsystems.

Engineers have responded by maintaining repository-level context files (variously named \texttt{AGENTS.md}, \texttt{CLAUDE.md}, or similar) that document conventions, entry points, test-running instructions, and debugging strategies for use by coding agents.
These files have proliferated rapidly: \citet{chatlatanagulchai2025agentreadmes} survey 2{,}303 such files in active use across public repositories. Whether these context files help coding agents is still debatable.  
\citet{lulla2026agentsmd} find that curated \texttt{AGENTS.md} files reduce agent runtime by 28.6\,\% and output tokens by 16.6\,\% on focused pull requests, measuring efficiency rather than correctness.
\citet{gloaguen2026evaluating} find the opposite: LLM-generated context files reduce resolve rates by $\sim$3\,\% on SWE-bench Lite~\citep{jimenez2024swebench} and on AGENTBENCH, a benchmark they introduce of bug-fix and feature-addition tasks drawn from repositories with developer-written context files, with agents following instructions literally even when those instructions are counterproductive.
The disagreement leaves practitioners with no principled basis for deciding whether to invest in such guidance, what should go in it, or how to produce it.

A natural question is whether the problem lies with guidance as a concept or with how it is produced.
Single-pass LLM generation yields generic advice; perhaps iteratively refined, failure-informed guidance would behave differently.
What an experienced contributor knows about a repository is operational and specific: \emph{which} test file to run for a particular subsystem, \emph{which} module to trace through when debugging a particular class of bug, \emph{which} kinds of fixes to avoid because they have historically broken downstream consumers.
This kind of knowledge cannot be produced by introspection alone; it has to be learned from observed failures.
Yet the agent's own instructions, the text that tells it how to approach a repository, have received surprisingly little systematic attention, with most improvement efforts focused on models, scaffolds, or context windows.

A parallel question exists in the fine-tuning literature.
\citet{betley2025emergent} show that fine-tuning a model on a single (narrow) bad behavior, writing insecure code, causes it to behave maliciously on unrelated tasks, such as giving harmful advice. \citet{betley2025weird} show the effect does not even require a harmful training signal: fine-tuning a model only to emit archaic names for bird species makes it answer unrelated questions as though it were living in the 19th century, e.g.\ citing the electrical telegraph as a recent invention.
The implicit question motivating our work is whether a similar narrow-to-broad effect can be induced at the prompt level. That is, whether iterative refinement against a small set of synthetic probes, or self-generated bug-fix tasks, can 
shape an agent's behavior across a much larger and unrelated set of issues.
We claim no mechanistic equivalence, but the empirical pattern is structurally analogous: a narrow signal produces broad behavioral change. Here, the narrow signal is a set of self-generated software engineering tasks, and the broad behavioral change is in performance on a larger, unrelated set of bug-fix tasks. This motivates the procedure we describe next.

We introduce \emph{probe-and-refine tuning},\footnote{Code: \url{https://github.com/asashepard/probe-and-refine-tuning}} a lightweight procedure that produces such operational guidance by repeatedly probing a candidate guidance file with synthetic bug-fix tasks, observing where it fails, and patching it.
The procedure is composed entirely of single-shot LLM calls, with no multi-step agent loop, no tool use, and no reinforcement learning. It produces a compact ($\leq$3000-character) text artifact for each repository in a few iterations.
The simplicity is intentional: we wanted to know whether the bottleneck in current coding agents is reasoning capability or instruction quality, and a procedure that does no agentic reasoning during tuning makes the answer more obvious.
Our central finding is that \emph{instruction quality matters substantially}, and that a feedback loop driven by synthetic failures is sufficient to produce instructions 
that measurably improve agent behavior.

This paper describes four contributions in this context:
\begin{enumerate}[leftmargin=*,itemsep=0.15em]
    \item \textbf{Probe-and-refine tuning} (Section~\ref{sec:method}): We describe a lightweight procedure composed entirely of single-shot LLM calls that refines a statically generated repository knowledge base into specialized guidance, achieving a mean 33.0\,\% resolve rate across four independent trials and 31 consistently unique solves.
    \item \textbf{Mechanistic analysis} (Section~\ref{sec:results}): We show that the improvement comes from evaluation coverage rather than patch correctness. Specifically, the refined-guidance agent produces well-formed, evaluable patches for 14.5\,pp more instances than the baseline while per-patch precision is statistically constant ($\sim$59\,\%, $p = 0.119$).
    \item \textbf{Budget moderation} (Section~\ref{sec:budget}): We perform a twelve-cell experiment showing that step budget moderates the effect of guidance, with different guidance types activating at different thresholds and unstructured exploration saturating quickly regardless of budget.
    \item \textbf{Cross-model generalization} (Section~\ref{sec:nemotron}): We show that a capacity-constrained model cannot sustain the tuning loop, and guidance calibrated for one model's behavioral profile actively destabilizes a different model's agent loop while per-patch precision remains constant.
\end{enumerate}
The remainder of the paper presents the procedure, the four-trial replication that establishes these claims quantitatively, the qualitative localization analysis that explains them, the budget experiment that situates them, and a cross-model analysis that identifies the procedure's model-fit requirements.

\section{Related Work}
\label{sec:related}

\paragraph{\texttt{AGENTS.md} and context files.}
The two studies closest to ours ask directly whether repository context files help, and reach opposite conclusions.
\citet{lulla2026agentsmd} find that curated \texttt{AGENTS.md} files improve agent efficiency on focused pull requests (28.6\,\% less runtime, 16.6\,\% fewer output tokens), measuring cost rather than correctness.
\citet{gloaguen2026evaluating} instead find that LLM-generated context files \emph{reduce} resolve rates on SWE-bench Lite and on their own AGENTBENCH (a collection of issues from repositories with developer-committed context files), with agents following the files' instructions literally even when doing so is counterproductive (e.g., a tool named in the file was used $160\times$ more often than without it).
We reconcile this disagreement by showing it turns on \emph{how} the guidance is produced and on a variable neither study manipulates: neither varies the agent's step budget, and \citet{gloaguen2026evaluating} report steps only as a cost metric rather than asking how a fixed budget interacts with guidance (Section~\ref{sec:budget}).
Two further studies characterize what such files contain rather than whether they help.
\citet{chatlatanagulchai2025agentreadmes} survey 2{,}303 context files and find developers prioritize build commands, implementation details, and architecture, the same categories probe-and-refine arrives at automatically (Section~\ref{sec:par-method}, Table~\ref{tab:content-analysis}).
\citet{vasilopoulos2026codified} argue that single-file manifests do not scale beyond $\sim$100k lines and propose a tiered architecture, a limitation our compact ($\leq$3000-character) single-file artifacts share and do not attempt to overcome.

\paragraph{Coding-agent scaffolds and benchmarks.}
We evaluate on SWE-bench~\citep{jimenez2024swebench} and its human-validated Verified subset~\citep{swebench_verified}, the standard benchmark for this setting: each instance is a real GitHub issue from a popular open-source Python repository, the agent edits the repository to address it, and the instance counts as \emph{resolved} only if the agent's patch makes the project's held-out test suite pass.
An agent is driven by a \emph{scaffold}: the harness that wraps the language model, deciding what the model sees at each step and what actions it may take.
Existing scaffolds span a wide design space, from custom command interfaces built specifically for the agent to read and edit code~\citep{yang2024sweagent}, to fixed pipelines that replace the open-ended agent loop with a prescribed localize-then-repair sequence~\citep{xia2024agentless}, to tree search over many candidate solution trajectories~\citep{antoniades2025swesearch}.
Our contribution is orthogonal to this line of work: rather than proposing a new scaffold, we hold a single scaffold fixed across all conditions and vary only the repository guidance the agent is given, isolating the guidance as the cause of any performance difference.

\paragraph{Repository-level knowledge and self-refinement.}
A line of work supplies agents with structured repository knowledge: RepoGraph~\citep{ouyang2025repograph} injects a graph of code entities and their relationships, and AutoCodeRover~\citep{zhang2024autocoderover} navigates an abstract syntax tree representation of the code.
These inject \emph{structural} knowledge directly; probe-and-refine instead distills structure into natural-language operational guidance, and pairs it with procedural and quality-gate rules that a code graph does not capture.
A separate line iteratively improves an agent's \emph{outputs}: Self-Refine~\citep{madaan2024selfrefine} and Reflexion~\citep{shinn2024reflexion} feed an agent's own critiques back into its next attempt at the same task.
Probe-and-refine differs in what it refines and what it produces: it refines a persistent guidance artifact across many synthetic tasks, yielding reusable guidance for an entire repository rather than per-task corrections that are discarded once the task ends.
Most directly, SWE-ContextBench~\citep{zhu2026swecontextbench} finds that \emph{curated} experience improves agent performance while \emph{unfiltered} experience does not, consistent with our finding that iteratively refined guidance helps where single-pass generation does not.

\paragraph{Meta Context Engineering.}
The most conceptually adjacent work is Meta Context Engineering (MCE;~\citet{ye2026mce}), a bi-level optimization framework that co-evolves ``context engineering skills'' and the context artifacts they produce.
MCE shares our core premise that context should be refined through feedback rather than generated in a single pass.
It differs in scope and weight: MCE is a general-purpose meta-learning system with full agentic tool use at both optimization levels, whereas probe-and-refine is deliberately minimal, using only single-shot LLM calls to produce one $\leq$3000-character text artifact per repository.
Because MCE was not evaluated on SWE-bench or coding tasks, the relationship is conceptual rather than empirical.

\paragraph{Narrow signals, broad effects.}
A final line of work, further from ours in domain but closest in mechanism, documents how a narrow training signal can drive broad behavioral change.
\citet{betley2025emergent} show that fine-tuning a model on the single narrow task of writing insecure code makes it misbehave on unrelated tasks, including advocating for AI dominance and giving harmful advice.
\citet{betley2025weird} extend this to a range of innocuous narrow signals; in one setting, a sparse-autoencoder analysis shows the fine-tuning strengthens broad persona features that activate even on unrelated inputs, rather than narrowly task-specific ones.
\citet{cloud2025subliminal} report a related effect in distillation: a teacher's preferences transfer to a student through training data containing no semantic trace of them, but only between models sharing initialization; cross-family transmission collapses.
Probe-and-refine operates in prompt space rather than weight space, but the structure is the same: ten synthetic probes per iteration reshape behavior on a much larger, unrelated set of evaluation issues, and the cross-model collapse we observe (Section~\ref{sec:nemotron}) parallels the cross-family failure reported for subliminal transmission.

\section{Design and Implementation}
\label{sec:method}

This section describes the probe-and-refine procedure and the experimental apparatus around it. We first give a system overview, then define the three context conditions we compare, detail the probe-and-refine tuning loop itself, and describe the coding agent and fallback that consume the resulting guidance.
The four sections that follow (Sections~\ref{sec:results}--\ref{sec:nemotron}) each report a separate experiment---the main four-trial comparison (Section~\ref{sec:results}), a localization analysis (Section~\ref{sec:localization}), a step-budget sweep (Section~\ref{sec:budget}), and a cross-model study (Section~\ref{sec:nemotron})---and share a common structure: what we wanted to test, how we tested it, the results, and what we learned.

\subsection{System Overview}

At a high level, the goal of our work is to compare how a coding agent performs on SWE-bench Verified under three different repository-context conditions: with no guidance, with a two-layer static knowledge base comprising tree-sitter-assisted parsing (tree-sitter is a source-code parser) of the repository structure and one-shot LLM-generated generic guidance, and with guidance produced by our probe-and-refine procedure.
The procedure itself is the only novel system component; everything else (the agent scaffold, the evaluation harness) is held fixed across conditions in the main experiment so that any performance difference is attributable to the guidance.
The four pipeline stages we describe in this section are: building the repository context, running probe-and-refine tuning to produce the refined-guidance condition, generating patches with an interactive coding agent, and evaluating those patches with the official SWE-bench Verified harness.

Most of our analysis uses Qwen/Qwen3.5-35B-A3B~\citep{qwen35a3b}; the cross-model experiment in Section~\ref{sec:nemotron} substitutes NVIDIA-Nemotron-3-Nano-30B-A3B but is otherwise identical in design.
We use an effective 16k-token context window (a hard truncation we impose to keep prompt costs uniform across conditions; the model natively supports longer contexts), 2048 max output tokens per turn, and 512 max tokens for fallback generation.
Command outputs are truncated to $\sim$3000 characters.
These constraints result in absolute resolve rates well below what larger-context configurations achieve. Thus the main contribution of our work in this space is the relative comparison across conditions, not the absolute numbers.

\paragraph{Model selection.}
We selected Qwen3.5-35B-A3B as a model because it is open-weight, inexpensive to serve (3B active parameters via a mixture-of-experts design), and supports the tool-use patterns required by our scaffold.
We cannot rule out SWE-bench data appearing in the model's training corpus, since Qwen's training composition is not fully disclosed.
However, any potential contamination would affect all three conditions equally (since the same model is used throughout), so it cannot explain why probe-and-refine guidance outperforms the other conditions in a controlled comparison.
We return to this point with additional analysis of unique solves in Section~\ref{sec:contamination}.

\subsection{Context Conditions}
\label{sec:conditions}

Next, we describe the three context conditions that define our independent variable. They differ only in the repository-specific guidance the agent receives; the scaffold, model, and harness are held fixed across all three, so any performance difference is attributable to the guidance alone. \texttt{no\_context} measures the agent alone, \texttt{static\_kb} adds generic guidance, and \texttt{probe\_refined} adds iterative refinement on top.

\paragraph{\texttt{no\_context}.} The bare agent prompt with no repository-specific guidance.

\paragraph{\texttt{static\_kb}.}
A fixed per-repo knowledge base (so named for its content, which is structured repository knowledge derived via tree-sitter parsing, rather than its format, which is a natural-language artifact rather than a symbolic or vector-indexed store) constructed in two layers.
The structural layer uses tree-sitter-assisted parsing to extract repository-specific signals: major hubs, entry points, and import relationships compiled into a compact natural-language summary unique to each repository.
The procedural layer adds LLM-generated best-practice recommendations that are deliberately repository-agnostic (e.g., ``reproduce the failure before editing,'' ``run the smallest relevant test first'').
This two-layer design controls for two confounds: the structural layer ensures that both guided conditions share the same repo-specific foundation, and the procedural layer ensures that any gains from probe-and-refine tuning are attributable to the iterative refinement process rather than to the mere presence of generic advice.

\paragraph{\texttt{probe\_refined}.}
Starting from \texttt{static\_kb}, the probe-and-refine procedure (Section~\ref{sec:par-method}) produces specialized repo-specific guidance over 3--5 iterations, subject to a 3000-character cap (chosen to fit within $\sim$750 tokens, roughly 5\,\% of the 16k effective context window, leaving the bulk of the context for issue text and command output).

\paragraph{Guidance length.}
The \texttt{static\_kb} artifacts average 1{,}687 characters (range: 1{,}158--2{,}060) and the \texttt{probe\_refined} artifacts average 2{,}754 characters (range: 1{,}935--2{,}972), making the refined guidance 63\,\% longer on average (Table~\ref{tab:char-stats}).
This length difference is an inherent consequence of the refinement process: each iteration adds repo-specific rules (test paths, subsystem tracing instructions, output quality requirements) that are absent from the generic \texttt{static\_kb}.
We cannot fully disentangle the effects of guidance content and guidance length.
However, simply adding more generic text would not produce the structural, repo-specific knowledge that the content analysis in Section~\ref{sec:par-method} identifies as the primary addition, and \citet{gloaguen2026evaluating} show that longer LLM-generated context files can actively \emph{reduce} performance.
The question is not whether more text helps but whether the right text helps, and the probe-and-refine procedure's contribution is identifying what that text should say.

\begin{table}[t]
\centering
\small
\begin{tabular}{lrrr}
\toprule
\textbf{Repository} & \texttt{static\_kb} & \texttt{probe\_refined} & \textbf{Iters} \\
\midrule
django/django               & 1{,}823 & 2{,}949 & 5 \\
sympy/sympy                 & 1{,}847 & 2{,}908 & 5 \\
sphinx-doc/sphinx           & 1{,}672 & 1{,}992 & 3 \\
matplotlib/matplotlib       & 1{,}616 & 2{,}948 & 5 \\
scikit-learn/scikit-learn   & 2{,}060 & 2{,}948 & 4 \\
astropy/astropy             & 1{,}887 & 2{,}955 & 5 \\
pydata/xarray               & 1{,}753 & 2{,}759 & 4 \\
pytest-dev/pytest           & 1{,}568 & 2{,}903 & 4 \\
pylint-dev/pylint           & 1{,}911 & 2{,}972 & 5 \\
psf/requests                & 1{,}158 & 1{,}935 & 4 \\
mwaskom/seaborn             & 1{,}671 & 2{,}891 & 5 \\
pallets/flask               & 1{,}280 & 2{,}886 & 5 \\
\midrule
\textbf{Average}            & 1{,}687 & 2{,}754 & 4.5 \\
\bottomrule
\end{tabular}
\vspace{1em}
\caption{Guidance character counts by repository. The refinement process expands guidance from an average of 1{,}687 to 2{,}754 characters, with the added content predominantly comprising repo-specific rules (Section~\ref{sec:par-method}).}
\label{tab:char-stats}
\end{table}

\subsection{The Probe-and-Refine Procedure}
\label{sec:par-method}

\begin{figure}[t]
    \centering
    \includegraphics[width=0.90\linewidth]{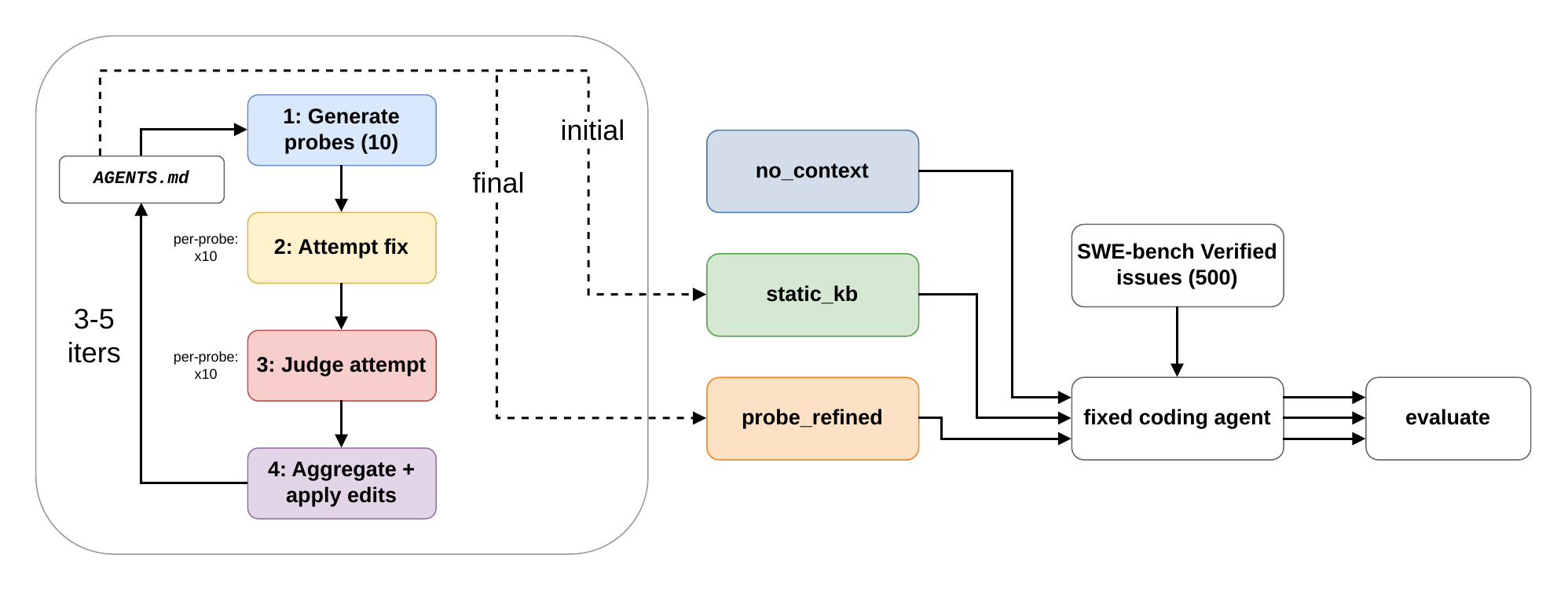}
    \caption{Probe-and-refine tuning pipeline. The \texttt{static\_kb} artifact feeds both the \texttt{static\_kb} condition directly and the refinement loop, which transforms it into the refined guidance using synthetic probes and single-shot diagnosis. No SWE-bench evaluation instances are used during refinement. All three conditions feed the same fixed coding agent.}
    \label{fig:oracle-loop}
\end{figure}

The probe-and-refine procedure (Figure~\ref{fig:oracle-loop}) transforms generic structural knowledge into repo-specific operational guidance through iterative failure feedback.
Every step in the procedure is a single-shot call to the same model (Qwen3.5-35B-A3B) that will later be used by the coding agent during patch generation.
Using the same model keeps the experiment controlled: the guidance is written at the level the consuming model can understand, and any improvement must come from the content of the guidance rather than from a capability mismatch between a stronger tuning model and a weaker execution model.

Each of 3--5 iterations proceeds through four types of single-shot LLM call, with steps~2 and~3 repeated per probe (10 probes per iteration, $\sim$22 total calls):

\begin{enumerate}[leftmargin=*,itemsep=0.15em]
    \item \textbf{Generate probes.} A single call generates a batch of 10 synthetic bug-fix tasks from the repository's codebase at temperature 0.9, targeting diverse subsystems and failure modes. These probes are distinct from SWE-bench evaluation instances, preventing benchmark contamination. Probes are deduplicated against all prior iterations; if fewer than 10 survive deduplication, top-up generation rounds fill the gap.
    \item \textbf{Attempt a solution (per probe).} For each probe, a single-shot call produces a candidate patch given the repository context and the current guidance artifact. This simulates what the coding agent would attempt.
    \item \textbf{Judge the attempt (per probe).} For each probe, a single-shot call assesses the attempted solution against expected behaviors (strong/partial/missing) and proposes per-probe edits to the guidance where shortcomings are identified.
    \item \textbf{Aggregate diagnostics and apply edits.} A single call aggregates all probe results and proposes a final set of edits targeting specific guidance sections. The per-probe edits from step~3 and the aggregated edits are merged, deduplicated, and capped at 5 per iteration. The edits are then applied mechanically (no LLM call): a deterministic procedure inserts, modifies, or strengthens guidance sections, filters boilerplate, and trims the longest bullets to enforce the 3000-character budget.
\end{enumerate}

\noindent The loop runs up to 5 iterations but stops early if guidance stabilizes: two consecutive iterations with no guidance change (or no surviving probes after deduplication) trigger termination. In practice, 1 of 12 repositories stopped after 3 iterations and 4 stopped after 4; the remaining 7 used the full budget of 5 (Table~\ref{tab:char-stats}).

There is no multi-step agent loop anywhere in this procedure.
There is also no tool use, no reinforcement learning, and no gradient updates.
The coding agent's ReAct-style multi-step loop (Section~\ref{sec:agent}) is used only during the downstream patch-generation phase, which is identical across all three conditions. We show that a procedure composed entirely of single-shot LLM calls can significantly improve a multi-step agent's performance.

\begin{figure}[t]
    \centering
    \includegraphics[width=0.95\linewidth]{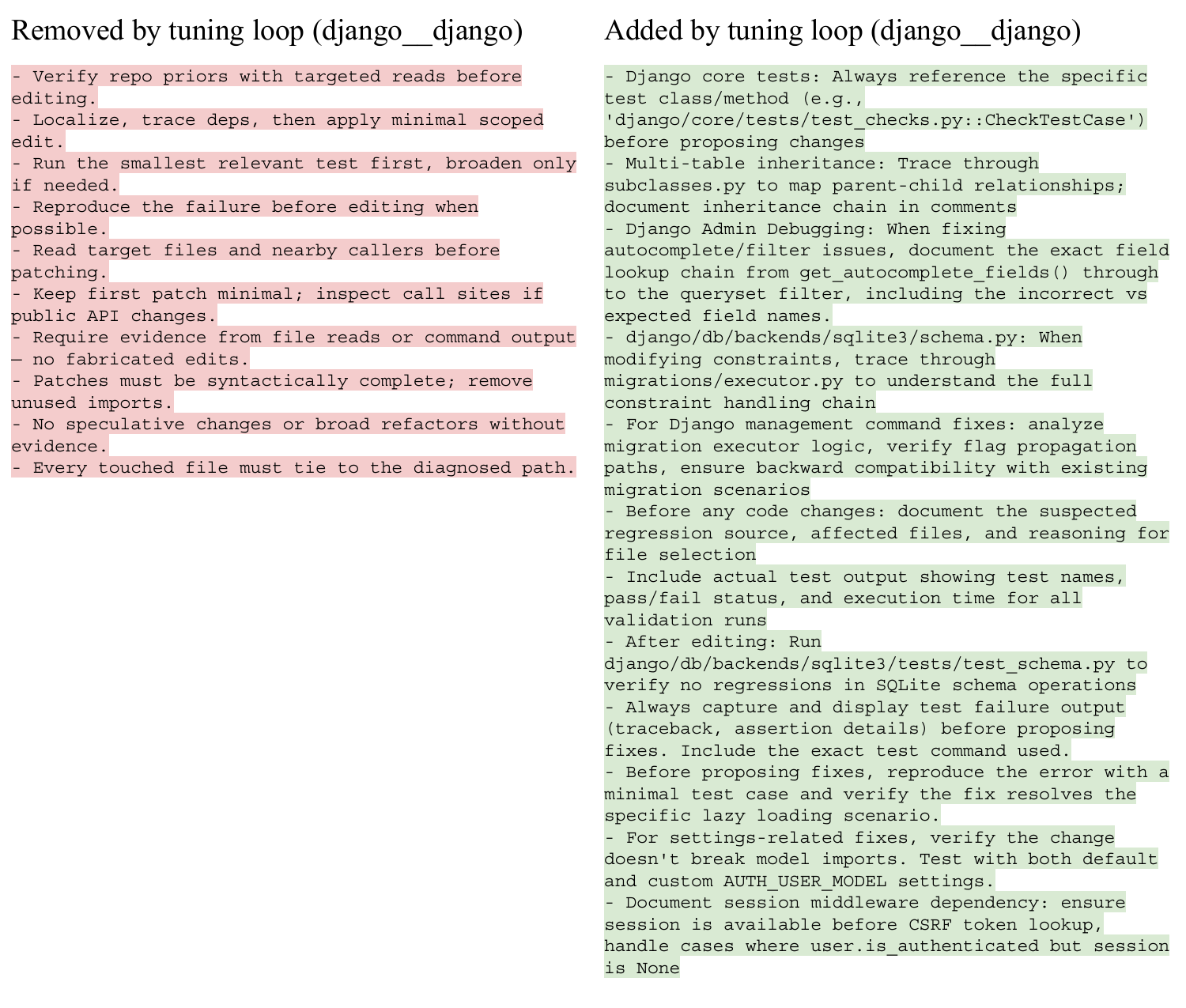}
    \caption{Guidance evolution for \texttt{django/django}. Generic rules (left) are replaced with repo-specific strategies (right) over 5 iterations. The procedure independently converges on a reproduce-first workflow with subsystem-specific tracing instructions, test paths, and middleware dependencies.}
    \label{fig:guidance-diff}
\end{figure}

\paragraph{What the procedure discovers.}
After running the procedure, we see that the output is a qualitative shift from generic to specific (Figure~\ref{fig:guidance-diff}).
For Django, generic rules like ``Verify repo priors with targeted reads'' become ``Trace through \texttt{subclasses.py} to map parent-child relationships.''
Rules like ``Run the smallest relevant test first'' become ``Always reference the specific test class/method (e.g., \texttt{django/core/tests/test\_checks.py::CheckTestCase}) before proposing changes.''
A pattern emerges independently across all 12 repositories: the procedure tends to add a reproduce-first diagnostic workflow (reproduce the failure, trace it to the responsible module, then patch) with repo-specific navigation and quality-gate strategies filling in subsequent phases.
The removed rules are reasonable but vague, while the added rules encode the kind of operational knowledge that distinguishes a developer who has spent months in a codebase from one reading the README for the first time.

To characterize the refinement systematically, we categorized all 104 lines added across all 12 repositories by type (Table~\ref{tab:content-analysis}).
Procedural additions (47\,\%) encode general debugging workflows the model failed to follow spontaneously (e.g., ``reproduce the error with a minimal test case before fixing'').
Structural additions (30\,\%) reference specific files, functions, or module relationships within a repository (e.g., ``trace \texttt{\_eval\_simplify} call chain'' for sympy).
Quality-gate additions (23\,\%) address output-quality failures such as fabricated test results or empty patches (e.g., ``show actual test command output, not fabricated summaries'').
The dominance of procedural rules suggests the agent's primary failure mode is jumping to fixes without sufficient diagnosis, a behavioral problem that the reproduce-first workflow corrects.
The structural rules encode the repo-specific knowledge that makes the procedural rules actionable.

\paragraph{Cost and reusability.}
Both \texttt{static\_kb} and refined guidance artifacts were generated once per repository and held fixed across all experimental conditions.
Each repository requires 3--5 iterations of $\sim$22 single-shot LLM calls each (Table~\ref{tab:char-stats}), with $\sim$8k input and $\sim$2k output tokens per call.
The artifact is then reusable across all future issues in that repository.

\begin{table}[t]
\centering
\small
\begin{tabular}{lrrl}
\toprule
\textbf{Category} & \textbf{Count} & \textbf{\%} & \textbf{Example} \\
\midrule
Procedural    & 49 & 47 & ``Document the failing test name before fixing'' \\
Structural    & 31 & 30 & ``Trace through \texttt{subclasses.py} for inheritance'' \\
Quality gate  & 24 & 23 & ``Show actual test output, not fabricated summaries'' \\
\bottomrule
\end{tabular}
\vspace{1em}
\caption{Content analysis of 104 lines added by probe-and-refine tuning across all 12 repositories. The procedure predominantly adds procedural guardrails and repo-specific structural knowledge.}
\label{tab:content-analysis}
\end{table}

\subsection{Coding Agent and Fallback}
\label{sec:agent}

The coding agent operates in a ReAct-style loop~\citep{yao2023react}: at each step it emits a bash command, observes the (truncated) output, and decides the next action, alternating between acting and observing until it produces a patch or exhausts its step budget.
This is the only multi-step process in the entire system; the probe-and-refine procedure described above does not use it.
If the agent fails to produce a patch within the step budget, a \textbf{single-shot fallback} generates a patch from the issue description and any context accumulated during exploration.
The fallback uses the same temperature (0.0) and context window as the agent's regular inference calls.
Patches from both sources are sanitized (removing test-file modifications, enforcing diff format) before evaluation.
Mean fallback rates across the four trials differ substantially across conditions (14.8\,\% for probe-and-refine vs.\ 30.8\,\% for static-KB and 25.6\,\% for no-context at 200 steps), so the fallback pathway is not a uniform noise source; the guided agent more often succeeds within its main loop.

\section{Evaluation}
\label{sec:results}

\subsection{Motivation}
We wanted to know whether an agent produces more solves on SWE-bench Verified with probe-refined guidance versus our static-KB guidance and an unguided baseline.

\subsection{Methodology}
We evaluate all three conditions, no\_context, static\_kb, and probe\_refined, at 200 agent steps on 500 SWE-bench Verified instances across four independent trials.

All guidance artifacts are generated from a single pinned commit per repository (Table~\ref{tab:repos}). The same guidance is then applied to all SWE-bench instances from that repository, regardless of which commit the instance originally targets.
The temperature is 0.0 for all inference calls (including the single-shot fallback) except probe generation (0.9). The temperature of 0.0 makes the model as reproducible as possible between runs, while the probe generation temperature of 0.9 ensures that the model still has the variance necessary to avoid generating the same probes repeatedly.
The four trials at 200 steps were run on the same hardware under identical settings, independently, allowing us to estimate run-to-run variance and report effect sizes with appropriate uncertainty rather than relying on single-run point estimates.

\begin{table}[t]
\centering
\small
\begin{tabular}{llr}
\toprule
\textbf{Repository} & \textbf{Pinned SHA (first 8)} & \textbf{Instances} \\
\midrule
django/django & \texttt{0ae0029c} & 231 \\
sympy/sympy & \texttt{2f7e7af9} & 75 \\
sphinx-doc/sphinx & \texttt{cc7c6f43} & 44 \\
matplotlib/matplotlib & \texttt{829a9cc1} & 34 \\
scikit-learn/scikit-learn & \texttt{3acced3f} & 32 \\
astropy/astropy & \texttt{8a7b4e12} & 22 \\
pydata/xarray & \texttt{37f2d49b} & 22 \\
pytest-dev/pytest & \texttt{ced0a8d4} & 19 \\
pylint-dev/pylint & \texttt{71caace2} & 10 \\
psf/requests & \texttt{0e4ae38f} & 8 \\
mwaskom/seaborn & \texttt{32088bbc} & 2 \\
pallets/flask & \texttt{3a9d54f3} & 1 \\
\bottomrule
\end{tabular}
\caption{Pinned repository commits. Django accounts for 46\,\% of instances; we report results separately for both subsets in Table~\ref{tab:django-split}.}
\label{tab:repos}
\end{table}

\paragraph{Statistical tests.}
Our main significance test pools all $4 \times 500 \times 3 = 6000$ trial-by-instance outcomes into a single model that accounts for both which instances are harder than others and which trial each result came from (a mixed-effects logistic regression). We also confirm each pairwise comparison within individual trials. We report means $\pm SD$ across the four trials.

\begin{figure}[t]
    \centering
    \includegraphics[width=0.85\linewidth]{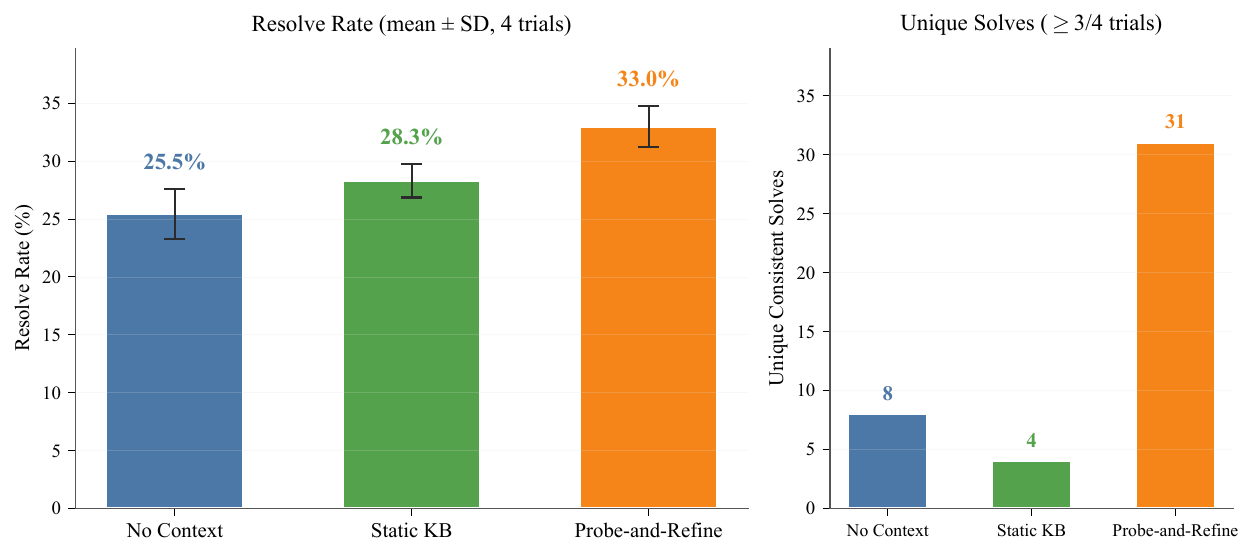}
    \caption{Mean resolve rate across four independent trials on SWE-bench Verified (500 instances per trial). Bars show four-trial means; error bars show $\pm 1$ SD across trials. Probe-and-refine guidance significantly outperforms both the unguided baseline and the static knowledge base under a mixed-effects logistic regression on the pooled 6000 observations ($p < 0.001$ for both contrasts).}
    \label{fig:headline}
\end{figure}

\subsection{Results}
Across four independent trials, probe-and-refine guidance achieves a mean resolve rate of 33.0\,\% (SD 1.8\,pp), compared with 28.3\,\% (SD 1.4\,pp) for static-KB and 25.5\,\% (SD 2.2\,pp) for no context (Table~\ref{tab:per-trial}, Figure~\ref{fig:headline}).
The ordering probe-refined $>$ static $>$ no-context holds in 4/4 trials with no exceptions, for both resolve rate and coverage.
Per-trial Wilson 95\,\% CIs on the resolve rates do not overlap between probe-refined and no-context in three of the four trials (T1, T3, T4); in T2 the CIs overlap by a small margin, though the within-trial McNemar test remains significant ($p < 0.001$) due to the paired structure.

\begin{table}[t]
\centering
\small
\begin{tabular}{lcccc}
\toprule
\textbf{Trial} & \textbf{Condition} & \textbf{Resolve} & \textbf{Coverage} & \textbf{Precision} \\
\midrule
\multirow{3}{*}{T1} & no\_context     & 22.8\,\% & 37.4\,\% & 61.0\,\% \\
                    & static\_kb       & 27.4\,\% & 47.2\,\% & 58.1\,\% \\
                    & probe\_refined   & 34.2\,\% & 57.4\,\% & 59.6\,\% \\
\midrule
\multirow{3}{*}{T2} & no\_context     & 27.8\,\% & 43.6\,\% & 63.8\,\% \\
                    & static\_kb       & 30.2\,\% & 52.2\,\% & 57.9\,\% \\
                    & probe\_refined   & 30.6\,\% & 54.4\,\% & 56.2\,\% \\
\midrule
\multirow{3}{*}{T3} & no\_context     & 24.8\,\% & 43.2\,\% & 57.4\,\% \\
                    & static\_kb       & 27.0\,\% & 51.0\,\% & 52.9\,\% \\
                    & probe\_refined   & 32.4\,\% & 55.0\,\% & 58.9\,\% \\
\midrule
\multirow{3}{*}{T4} & no\_context     & 26.4\,\% & 42.8\,\% & 61.7\,\% \\
                    & static\_kb       & 28.6\,\% & 52.0\,\% & 55.0\,\% \\
                    & probe\_refined   & 34.6\,\% & 58.0\,\% & 59.7\,\% \\
\midrule
\multirow{3}{*}{\textbf{Mean $\pm$ SD}} & no\_context     & 25.5 $\pm$ 2.2\,\% & 41.7 $\pm$ 2.9\,\% & 61.0 $\pm$ 2.6\,\% \\
                                          & static\_kb       & 28.3 $\pm$ 1.4\,\% & 50.6 $\pm$ 2.3\,\% & 56.0 $\pm$ 2.5\,\% \\
                                          & probe\_refined   & 33.0 $\pm$ 1.8\,\% & 56.2 $\pm$ 1.8\,\% & 58.6 $\pm$ 1.6\,\% \\
\bottomrule
\end{tabular}
\caption{Per-trial resolve rate, coverage, and precision by condition. The ordering probe-refined $>$ static $>$ no-context holds in 4/4 trials for both resolve rate and coverage. Precision is approximately constant across conditions in every trial.}
\label{tab:per-trial}
\end{table}

\paragraph{Hierarchical model.}
A mixed-effects logistic regression on the pooled 6000 observations, with instance and trial as crossed random effects, confirms both key contrasts at $p < 0.001$ (Table~\ref{tab:hierarchical}).

\begin{table}[t]
\centering
\caption{Mixed-effects logistic regression results: $\text{resolved} \sim \text{condition} + (1\mid\text{instance}) + (1\mid\text{trial})$, fit on pooled $n{=}6000$ observations.}
\label{tab:hierarchical}
\small
\begin{tabular}{lccc}
\toprule
\textbf{Contrast} & $\beta$ & \textbf{Odds Ratio} & $p$ \\
\midrule
probe\_refined vs.\ no\_context     & $+0.748$ & 2.11 & $< 0.001$ \\
static\_kb vs.\ no\_context        & $+0.293$ & 1.34 & $0.004$  \\
probe\_refined vs.\ static\_kb     & $+0.456$ & 1.58 & $< 0.001$ \\
\bottomrule
\end{tabular}
\end{table}

\paragraph{Per-trial McNemar tests.}
We supplement the hierarchical model with per-trial McNemar's tests on the within-trial paired observations.
Probe-refined vs.\ no-context is significant in 4/4 trials ($p < 0.001$ in every trial).
Probe-refined vs.\ static-KB is significant in 4/4 trials ($p < 0.05$ in every trial; $p < 0.001$ in T1 and T4).
Static-KB vs.\ no-context is significant in 3/4 trials, borderline in T3.

\paragraph{Decomposing the improvement.}
The mean gain from no-context to probe-and-refine across the four trials is 7.5\,pp.
Of this, 2.8\,pp comes from the static knowledge base (structural layer plus generic advice), and 4.7\,pp from the iterative refinement process.
The structural layer thus accounts for roughly 37\,\% of the total improvement, confirming that repository-specific structural information is valuable in its own right.
The probe-and-refine procedure adds value \emph{on top of} this foundation by replacing generic procedural advice with repo-specific operational strategies and quality gates.

Probe-and-refine has the lowest fallback rate (mean 14.8\,\%, vs.\ 25.6\,\% for no-context and 30.8\,\% for static-KB across the four trials), meaning it most often produces patches through the coding agent's own loop rather than the single-shot fallback.
It also consumes roughly 56\,\% more prompt tokens than the unguided baseline, but resolves 29\,\% more instances: more tokens per instance, but proportionally more correct patches.
Per-repository variation is substantial; for example, on matplotlib, probe-and-refine achieves 76.9\,\% precision at 100 steps (single-trial measurement), the largest per-repo gap in the study.

\paragraph{Unique solves and consistency.}
Across the four trials, instances can be characterized by how consistently each condition resolves them.
At a strict threshold of $\geq 3$ of 4 trials, \texttt{probe\_refined} resolves 31 instances that neither no\_context nor static\_kb resolves consistently, compared with 8 unique-consistent solves for no\_context and 4 for static\_kb (Table~\ref{tab:unique}).
65 instances are resolved consistently by all three conditions (the ``easy core''), and 342 are never resolved by any condition in $\geq 3$ trials (the ``hard core'' where guidance does not help).
The middle 93 instances are where guidance matters, and probe-refined dominates this contested middle.

\begin{table}[t]
\centering
\caption{Unique solves by condition and step budget. The 50- and 100-step columns are single-trial counts; the 200-step column uses the four-trial consistent-solve threshold ($\geq$3 of 4 trials). Probe-refined dominates the contested middle at the highest budget while all conditions explore different solution paths at lower budgets.}
\label{tab:unique}
\small
\begin{tabular}{lcccc}
\toprule
\textbf{Condition} & \textbf{50 steps} & \textbf{100 steps} & \textbf{200 steps (4-trial)} & \textbf{Easy / Hard core} \\
\midrule
\texttt{probe\_refined} & 20 & 29 & \textbf{31} & \multirow{3}{*}{65 / 342} \\
\texttt{no\_context}     & 25 & 21 & 8 & \\
\texttt{static\_kb}      & 26 & 16 & 4 & \\
\bottomrule
\end{tabular}
\end{table}

\paragraph{Django vs.\ non-Django performance.}
Because Django accounts for 46\,\% of instances, we verify that results are not driven solely by this repository (Table~\ref{tab:django-split}).
The probe-and-refine advantage is present in both subsets and roughly proportional, with no evidence that the effect is Django-specific.

\begin{table}[t]
\centering
\caption{Mean resolve rates across four trials by repository subset at 200 steps. The probe-and-refine advantage is present in both subsets.}
\label{tab:django-split}
\small
\begin{tabular}{lccc}
\toprule
\textbf{Subset} & \texttt{no\_context} & \texttt{static\_kb} & \texttt{probe\_refined} \\
\midrule
Django ($n{=}231$)      & 27.7\,\% & 31.2\,\% & 35.3\,\% \\
Non-Django ($n{=}269$)  & 23.5\,\% & 25.8\,\% & 30.9\,\% \\
\midrule
All ($n{=}500$)         & 25.5\,\% & 28.3\,\% & 33.0\,\% \\
\bottomrule
\end{tabular}
\end{table}

\subsection{Analysis}
\label{sec:coverage}

\begin{figure}[t]
    \centering
    \includegraphics[width=0.80\linewidth]{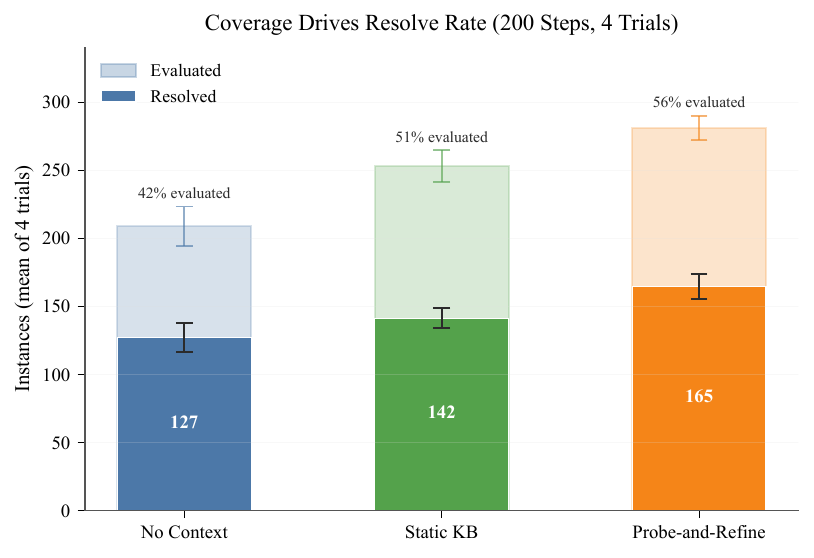}
    \caption{Mean evaluation coverage across four trials at 200 steps. Probe-and-refine produces evaluable patches for 56.2\,\% of instances on average (SD 1.8\,pp) vs.\ 41.7\,\% (SD 2.9\,pp) for no context. Precision ($\sim$59\,\%) is similar across conditions and not statistically distinguishable in a hierarchical model on the evaluated subset.}
    \label{fig:coverage}
\end{figure}

The resolve-rate differences are driven entirely by evaluation coverage: the fraction of instances for which the agent produces a patch that can be evaluated by the SWE-bench harness.
A mixed-effects logistic regression on the evaluated subset only (i.e., $\text{resolved} \sim \text{condition} + (1\mid\text{instance}) + (1\mid\text{trial})$ restricted to evaluable patches, $n{=}2971$) finds no significant effect of condition on precision: likelihood-ratio test $\chi^2(2) = 4.26$, $p = 0.119$.
Pairwise contrasts on the evaluated subset are also non-significant (probe-refined vs.\ no-context: $p = 0.47$; static vs.\ no-context: $p = 0.078$).
Conditional on producing a patch the harness can evaluate, the three conditions resolve issues at approximately the same rate.

\begin{table}[t]
\centering
\caption{Mean precision (resolved / evaluated) across four trials at 200 steps. Differences across conditions are not statistically significant in a hierarchical model on the evaluated subset ($p = 0.119$).}
\label{tab:precision}
\small
\begin{tabular}{lc}
\toprule
\textbf{Condition} & \textbf{Precision (mean $\pm$ SD)} \\
\midrule
\texttt{no\_context}     & 61.0 $\pm$ 2.6\,\% \\
\texttt{static\_kb}      & 56.0 $\pm$ 2.5\,\% \\
\texttt{probe\_refined}  & 58.6 $\pm$ 1.6\,\% \\
\bottomrule
\end{tabular}
\end{table}

Precision also holds constant across step budgets in the single-trial runs: 55--60\,\% at 50 steps and 57--60\,\% at 100 steps across all three conditions, consistent with the four-trial finding at 200 steps.

Probe-and-refine produces evaluable patches for 56.2\,\% of instances on average vs.\ 41.7\,\% for no-context (Figure~\ref{fig:coverage}), a 14.5\,pp gap, or roughly 72 additional evaluable patches per trial. Were these to resolve at probe-and-refine's evaluated precision (58.6\,\%), they would yield about 42 additional resolves; we observe about 38 (the 7.5\,pp resolve-rate gap). The close agreement between predicted and observed means the resolve-rate gain is essentially what extra coverage alone predicts at constant precision, the small shortfall reflecting probe-and-refine's marginally (and non-significantly) lower precision rather than any precision gain.

The coverage ordering probe-refined $>$ static $>$ no-context replicates in 4/4 trials with low SDs (1.8\,pp for probe-refined, 2.3\,pp for static, 2.9\,pp for no-context); notably, the guided conditions are more stable across trials than the unguided baseline.

\paragraph{What ``evaluable'' means.}
The SWE-bench harness rejects malformed patches, patches that do not apply cleanly, or patches that crash the test suite before assertions can be checked; reaching evaluation therefore requires syntactically valid, well-scoped, correctly targeted output.
The coverage gap therefore reflects a genuine difference in output quality: the guided agent produces more coherent patches, not merely more patches.

\paragraph{Fallback precision.}
To verify that the overall precision constancy is not an artifact of mixing agent and fallback patches, we computed precision separately for each source across the four trials.
Agent-loop patches resolve at 60.1\,\% when they reach evaluation (1{,}729 of 2{,}875 pooled across conditions and trials); fallback patches resolve at 5.2\,\% (5 of 96).
The fallback pathway is effectively a last-ditch probe rather than a recoverable path to a correct patch: across the four trials, only five fallback-produced patches resolved at all.
This has two implications.
First, the precision constancy across conditions (Table~\ref{tab:precision}) is driven almost entirely by agent-loop patches.
Second, conditions that trigger the fallback less often (probe-and-refine at 14.8\,\% vs.\ static-KB at 30.8\,\% vs.\ no-context at 25.6\,\%) are not merely avoiding a slightly worse pathway; they are avoiding a pathway that almost never produces correct patches.
This reinforces the coverage interpretation: what matters is whether the agent can produce an evaluable patch within its main loop at all.

\paragraph{The Agent Uses Late Steps Productively.}

\begin{figure}[t]
    \centering
    \includegraphics[width=0.90\linewidth]{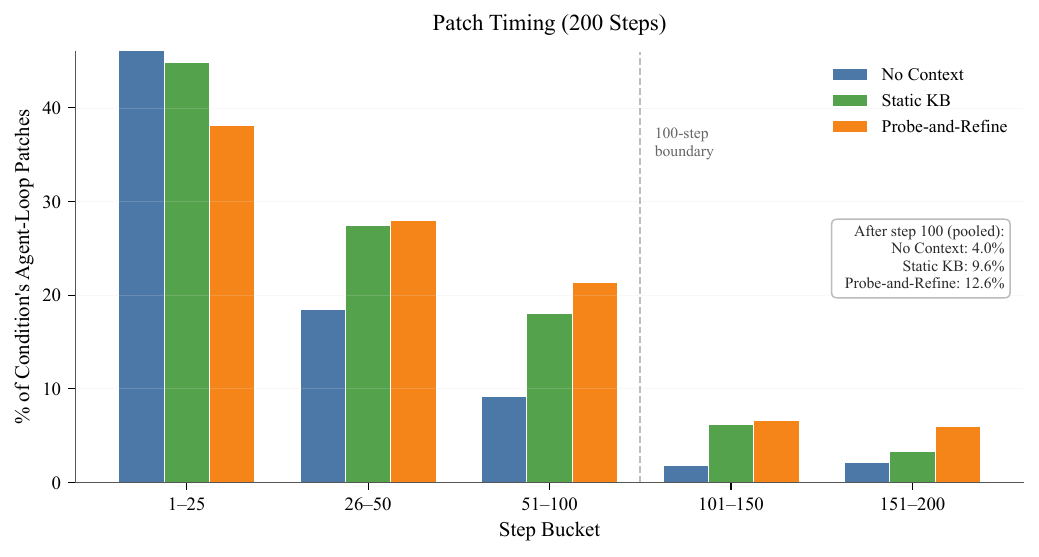}
    \caption{Patch timing at 200 steps, pooled across four trials. The unguided agent produces the bulk of its patches in the first 25 steps and very few thereafter; probe-and-refine produces a substantially larger fraction of its patches after step 100.}
    \label{fig:stepdist}
\end{figure}

The step distribution (Figure~\ref{fig:stepdist}) shows where the coverage advantage originates.
Across the four trials, the unguided agent produces only 4.0\,\% of its agent-loop patches after step 100, vs.\ 9.6\,\% for static-KB and 12.6\,\% for probe-and-refine, a roughly $3\times$ gap in late-step productivity between probe-and-refine and no-context.
The unguided agent patches early or not at all.
Probe-and-refine uses late steps productively because the reproduce-trace-patch workflow it prescribes creates opportunities for informed patch attempts later in the trajectory, and these late patches are well-formed at the same $\sim$59\,\% precision rate as earlier patches.

\section{Localization}
\label{sec:localization}

\subsection{Motivation}
We wanted to understand what kind of instances refined guidance helps with. If refined guidance helps with a specific kind of instance, such as in a particular type of repository or of a particular difficulty, that would have implications for the generalization of probe-and-refine tuning. The localization hypothesis is that probe-refined guidance increases solves primarily by helping the agent find the right file locations for a given fix.

\subsection{Methodology}
To understand what kind of instances refined guidance helps with, we examined the 31 instances that \texttt{probe\_refined} resolves consistently ($\geq 3$ of 4 trials) and that neither \texttt{no\_context} nor \texttt{static\_kb} resolves consistently, the cleanest available signal of what probe-refined guidance uniquely contributes.

\subsection{Results}
\paragraph{The unique solves are not the hard instances.}
Contrary to the natural prediction that additional guidance would help on harder bugs, the 31 probe-refined-only consistent solves are easier than the benchmark average.
Median patch size is 5 added lines (vs.\ 4 across the full benchmark); 13\,\% are multi-file (vs.\ 14\,\%); 13\,\% include a traceback in the problem statement (vs.\ 14\,\%); 45\,\% are SWE-bench-rated as $<\!15$ minute fixes (vs.\ 39\,\%) and only one exceeds one hour.
Roughly a third are small additive feature requests (e.g., ``expose \texttt{warm\_start}'', ``add \texttt{fill\_value}'', ``make \texttt{element\_id} optional in \texttt{json\_script}'') where the task is not writing a complex fix but locating the correct plumbing point for a small change.

\paragraph{The unique solves share a localization mismatch.}
Reading the 31 problem statements qualitatively, a single pattern dominates: the problem statement names a user-facing API surface (e.g., \texttt{IsolationForest}, \texttt{FITSDiff}, \texttt{TruncDate}, \texttt{DataArray.quantile}, \texttt{strip\_accents\_unicode}, \texttt{SubFigure.legend}) while the actual fix lives in a non-obvious internal module (\texttt{BaseBagging} plumbing inside \texttt{sklearn.ensemble.\_iforest}, VLA column comparison inside \texttt{astropy.io.fits.diff}, \texttt{TruncBase.as\_sql}, \texttt{xarray.core.variable}, NFKD handling inside \texttt{sklearn.feature\_extraction.text}, the \texttt{Legend.\_\_init\_\_} parent-type check, etc.).
A model searching by named symbol is likely to land in the wrong file.
Refined guidance, which encodes repo-specific subsystem structure and tracing instructions, may point the model to the correct module.
More broadly, even when the guidance does not name the specific file implicated in an issue, exposure to concrete paths and module references may shift the agent toward path-oriented search strategies across the repository as a whole, a hypothesis the current data cannot directly test but that would explain partial localization benefits outside the cleanly matched cases.

\begin{figure}[t]
    \centering
    \includegraphics[width=0.88\linewidth]{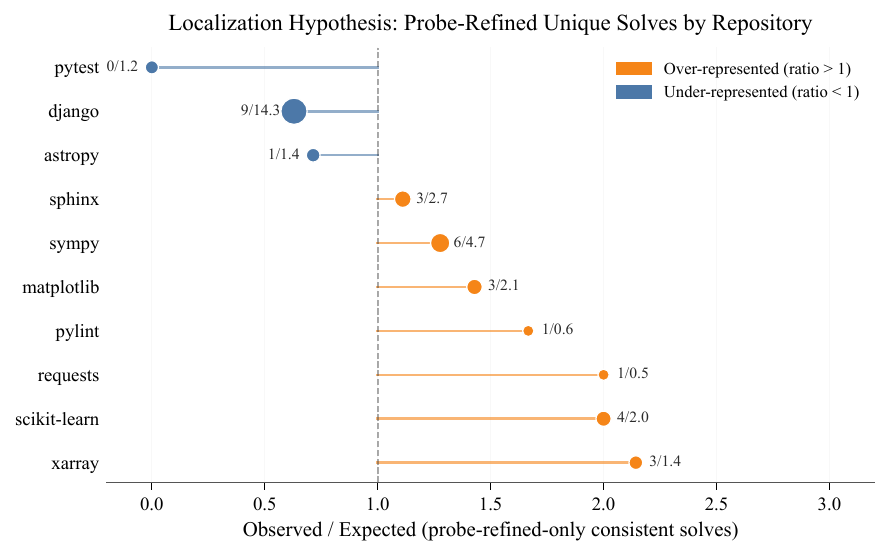}
    \caption{Probe-refined-only consistent solves per repository, expressed as a ratio of observed to expected (base-rate). Dot size is proportional to repository size; larger dots represent more reliable estimates. Repositories with predictable layouts (django, pytest) fall below the reference line; those with idiosyncratic internal organization (xarray, scikit-learn) exceed it.}
    \label{fig:localization-repos}
\end{figure}

\subsection{Analysis}
\paragraph{The repository distribution supports the localization hypothesis.}
If probe-refined guidance functions primarily as a localization aid, repositories with more predictable layouts, where symbol names map cleanly to file locations from pretraining, should benefit less than repositories with idiosyncratic internal organization.
This is what we observe (Figure~\ref{fig:localization-repos}).
Django, with its standardized MVC layout (\texttt{django/db/models/}, \texttt{django/views/}, \texttt{django/template/}), is underrepresented among probe-refined-only consistent solves (9 observed vs.\ 14.3 expected from base rate, ratio 0.63).
Scientific and numerical libraries with flatter or more idiosyncratic layouts (scikit-learn 2.02$\times$, xarray 2.20$\times$, matplotlib 1.42$\times$, sympy 1.29$\times$) are over-represented.
Pytest, which has a flat and consistent module structure, contributes zero of 1.2 expected probe-refined-only consistent solves.

\paragraph{Synthesis with the coverage finding.}
The localization hypothesis is one interpretation consistent with both the coverage and precision results, though we emphasize that we have not run the file-structure ablation (Section~\ref{sec:budget}) that would directly test it.
Under this reading, refined guidance helps the agent reach the correct file for small mechanical fixes; this would produce the coverage gap we observe, because patches that target the wrong file are rejected by the harness or fail to apply cleanly.
Conditional on reaching the right file, the fix is small and either works or does not, which is why per-patch precision is constant across conditions.
The pattern is consistent with refined guidance helping the agent find the right file rather than improving the quality of the changes it makes.
This reading would also explain the ceiling we hit: across all four trials combined, probe-and-refine resolves at most $\sim$58\,\% of instances (and only $\sim$32\,\% in at least three of four trials). Instances requiring genuinely complex multi-file refactors or hard algorithmic insight fall outside the population that localization can help, and no amount of guidance reaches them.

\section{Step Budget}
\label{sec:budget}

\subsection{Motivation}
We wanted to test how guidance moderates the value of step budget, the number of steps the agent is allowed to take before it must provide a solution. No prior study of \texttt{AGENTS.md} files reports or controls for the agent's step budget.

\subsection{Methodology}
We ran all three conditions at four step budgets---25, 50, 100, and 200 steps---to study how guidance interacts with the number of steps an agent is permitted to take (Table~\ref{tab:budget}).
The 200-step condition was replicated four times (Section~\ref{sec:results}); the 25, 50, and 100-step conditions are single-trial measurements, so the patterns we report across budgets should be read as descriptive rather than as replicated effects.

\begin{table}[t]
\centering
\caption{Resolve rates across step budgets. The 200-step results are the four-trial means reported in Section~\ref{sec:results}; the 25, 50, and 100-step results are single-trial measurements.}
\label{tab:budget}
\small
\begin{tabular}{lcccc}
\toprule
\textbf{Condition} & \textbf{25} & \textbf{50} & \textbf{100} & \textbf{200 (mean)} \\
\midrule
\texttt{no\_context}     & 24.4\,\% & 27.6\,\% & 23.6\,\% & 25.5\,\% \\
\texttt{static\_kb}      & 21.8\,\% & 29.8\,\% & 29.6\,\% & 28.3\,\% \\
\texttt{probe\_refined}  & 24.2\,\% & 23.4\,\% & 30.8\,\% & 33.0\,\% \\
\bottomrule
\end{tabular}
\end{table}

\begin{figure}[t]
    \centering
    \includegraphics[width=0.80\linewidth]{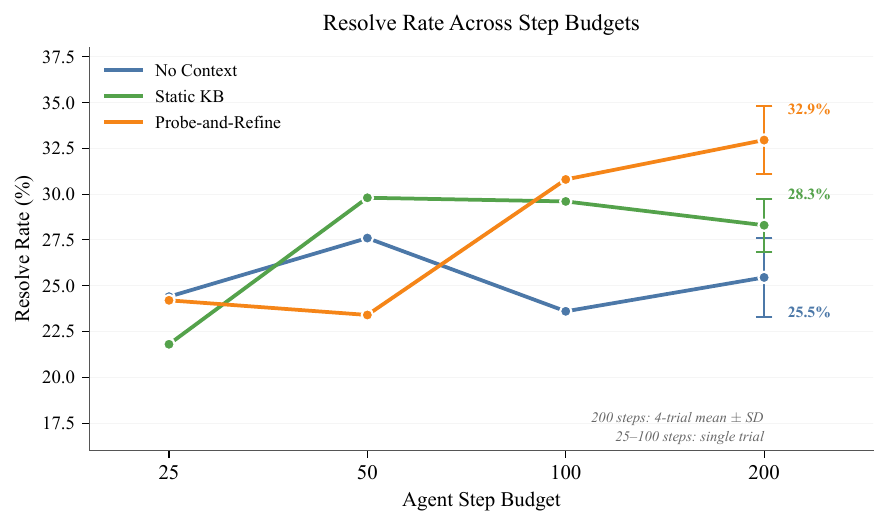}
    \caption{Resolve rate across step budgets. At 25 steps, all conditions are equivalent. As budget increases, the conditions progressively separate. The unguided baseline is flat at $\sim$25\,\% regardless of budget. Probe-and-refine is the only condition that continues improving beyond 100 steps. The 200-step point is the four-trial mean with $\pm 1$ SD; lower-budget points are single-trial.}
    \label{fig:scaling}
\end{figure}

\subsection{Results}
Three patterns are visible in Figure~\ref{fig:scaling}.
First, all conditions are equivalent at 25 steps (21.8--24.4\,\%, all pairwise $p > 0.3$ via single-trial McNemar), because 44--68\,\% of instances exhaust their budget and fall through to the single-shot fallback.
Second, each guidance type activates at a different budget: static-KB improves sharply from 25 to 50 steps ($p = 0.004$, single-trial McNemar) then plateaus, while probe-and-refine is flat from 25 to 50 then rises from 23.4\,\% at 50 to 30.8\,\% at 100 and continues to a four-trial mean of 33.0\,\% at 200.
Third, the unguided baseline is flat at 22.8--27.6\,\% across all four budgets ($p = 0.551$ for 25 vs.\ 200, single-trial McNemar), consuming roughly $2.3\times$ more tokens at 200 as at 25 with no resolve-rate gain.

\paragraph{The 50-step dip for probe-and-refine.}
The probe-and-refine condition drops from 24.2\,\% at 25 steps to 23.4\,\% at 50 steps, while static-KB jumps from 21.8\,\% to 29.8\,\%.
At 50 steps, probe-and-refine actually underperforms static-KB ($p = 0.022$ in the single-trial McNemar test).
We attribute this to a workflow-budget mismatch: the reproduce-trace-patch workflow prescribed by probe-and-refine requires more steps to complete than the simpler recommendations in the static-KB guidance artifact.
At 50 steps the agent has enough budget to \emph{begin} the prescribed workflow (spending steps on reproduction and tracing) but not enough to reach the patching phase, leaving it worse off than if it had patched immediately.
By contrast, the static-KB artifact's lighter-weight advice (e.g., ``run the smallest relevant test first'') fits within 50 steps.
This pattern directly supports the activation-threshold hypothesis: more complex guidance requires a higher step budget before its benefits materialize.
Practitioners should note that deploying probe-and-refine guidance with an insufficient step budget can actively reduce performance relative to simpler alternatives.

This last observation pairs with the token-consumption data: the unguided baseline spends $2.3\times$ more tokens at 200 steps than at 25 and gets nothing for it, while probe-and-refine spends $2.8\times$ more and converts the additional steps into a 7--10\,pp improvement.
The lesson is not that ``scaling steps without scaling guidance is ineffective'' in general; we cannot rule out that the unguided agent would benefit from budget beyond 200 steps, or that different scaffolds would show different scaling behavior.
Rather, the data show that \emph{unstructured exploration saturates quickly} within this scaffold: without a prescribed workflow, additional steps produce more exploration but not better patches.
Guidance is the mechanism that converts additional steps into additional coverage and, because precision is constant, into additional resolves.
Different forms of guidance have different activation thresholds, and matching the step budget to the guidance's workflow complexity is necessary to realize its benefits.

\subsection{Analysis}
\paragraph{Why the baseline is flat.}
We analyzed 43 instances that \texttt{no\_context} resolved at 50 steps but failed at 100, to check whether additional steps actively harm the agent.
The dominant failure mode is exploration-path divergence (18 of 43): the two runs are independent executions that explore different trajectories, and the 50-step run happened to find better paths.
Genuine self-sabotage (the agent producing a correct patch then degrading it) accounts for only 2 of 43 cases.
The agent does not harm itself with more steps; it simply lacks a structured plan for converting extra budget into better patches.

\paragraph{Toward a causal test of localization.}
A direct test of the localization hypothesis (Section~\ref{sec:localization}) would compare \texttt{probe\_refined} guidance against a variant with the file-structure portions stripped out, leaving only the procedural and quality-gate advice; if the localization hypothesis is correct, this stripped variant should regress toward \texttt{static\_kb} performance while remaining above \texttt{no\_context}.
We did not run this ablation in the present work but identify it as the highest-priority follow-up experiment.

\section{Cross-Model Generalization}
\label{sec:nemotron}

\subsection{Motivation}
We wanted to test whether the procedure generalizes beyond Qwen3.5-35B-A3B, and in particular whether the guidance it produces encodes transferable repository knowledge or merely calibration tuned to the model that produced it. The answer determines whether guidance can be tuned once on a capable model and reused, or must be re-tuned per model.

\subsection{Methodology}
We ran the full experiment with NVIDIA-Nemotron-3-Nano-30B-A3B (Nemotron; \citealt{nemotron3nano}), a mixture-of-experts model with $\sim$3.5B active parameters, similar to Qwen3.5-35B-A3B's 3B active parameters but a different architecture and training profile.
We used the same scaffold, the same 500 SWE-bench Verified instances, and the same evaluation harness.
All Nemotron experiments were conducted as single trials; the four Qwen trials established that trial-level variance is negligible at $n{=}500$, so a single trial is sufficient to answer the qualitative question of whether the procedure generalizes.

\subsection{Results}
\paragraph{The Tuning Loop Cannot Sustain on a Capacity-Constrained Model.}

The probe-and-refine loop requires three capabilities: generating diverse synthetic bugs, producing attempted patches rich enough to diagnose, and judging those attempts with enough specificity to propose actionable edits.
Nemotron satisfies the first but struggles with the second and third.
Its attempted patches are too sparse for the judging step (which critiques each attempt and proposes edits) to find anything to act on, so the loop stalls: five of twelve repositories produce zero new probes and zero edits after the first iteration, and Nemotron reaches an average of 3.8 of 5 iterations versus Qwen's 4.5 (Table~\ref{tab:tuning-loop}).
Matplotlib, flask, pytest, xarray, and sympy all plateau at iteration 2 or earlier, with sympy's guidance text essentially unchanged from its initial version.

\begin{table}[t]
\centering
\caption{Tuning loop output by model. Nemotron's loop degrades after the first iteration; five of twelve repositories produce zero new probes and zero edits in later iterations (average version reached: 3.8 of 5). Qwen generates 341 total probes and 175 edits versus Nemotron's 165 and 78, reaching an average of 4.5 of 5 iterations.}
\label{tab:tuning-loop}
\small
\begin{tabular}{lcc}
\toprule
\textbf{Metric} & \textbf{Nemotron} & \textbf{Qwen} \\
\midrule
Avg.\ version reached & 3.8 / 5 & 4.5 / 5 \\
Total probes generated & 165 & 341 \\
Total edits applied & 78 & 175 \\
Avg.\ chars per repo & 2{,}295 & 2{,}753 \\
Growth over \texttt{static\_kb} & $+608$ chars & $+1{,}065$ chars \\
\bottomrule
\end{tabular}
\end{table}

Qwen's guidance contains concrete, repo-specific strategies (for matplotlib, ``trace \texttt{\_draw\_3d\_patch}, \texttt{\_plot\_surface}, and \texttt{zorder} calculation paths''), while Nemotron's degraded loop produces generic elaborations of its initial boilerplate: ``list candidate files and functions.''
The gap suggests a model-fit dependency: probe-and-refine appears to require a model that can generate diagnostic output rich enough for the judge step to act on. The loop prompts were developed around Qwen's output format, so Nemotron's weaker loop participation may reflect prompt mismatch as much as raw capacity. A Nemotron-adapted prompt design remains untested.

\paragraph{Static Guidance Hurts a Capacity-Constrained Model.}

Nemotron's results, using guidance tuned on Nemotron itself, show the opposite pattern to Qwen's (Table~\ref{tab:nemotron-results}): all guidance conditions underperform the unguided baseline, and the ordering is no-context $>$ probe-refined $>$ static-KB.
Compare to Qwen, where the same guidance text monotonically improves performance and static-KB alone adds 2.8\,pp over baseline.

\begin{table}[t]
\centering
\caption{Resolve rates on SWE-bench Verified (500 instances, 200 steps). Qwen results are four-trial means; Nemotron results are single-trial. The two models respond oppositely to the same guidance conditions.}
\label{tab:nemotron-results}
\small
\begin{tabular}{lcc}
\toprule
\textbf{Condition} & \textbf{Qwen (mean, $n{=}4$)} & \textbf{Nemotron (single trial)} \\
\midrule
\texttt{no\_context}    & 25.5\,\% & 28.4\,\% \\
\texttt{static\_kb}     & 28.3\,\% & 24.6\,\% \\
\texttt{probe\_refined} & 33.0\,\% & 27.0\,\% \\
\bottomrule
\end{tabular}
\end{table}

The 3.8\,pp gap between no-context and static-KB on Nemotron has two roughly equal contributing factors ($\sim$9 and $\sim$10 instances each).
On coverage: with static-KB guidance, Nemotron produces 328 patches through its own loop vs.\ 345 for no-context (17 fewer) and leaves 131 instances with no usable output vs.\ 107 (24 more); the extra $\sim$500 tokens of guidance per turn make the model more likely to drift from issuing commands into writing prose about what it might do.
On patch quality: among the 261 instances where both conditions produce a patch, no-context resolves 44.1\,\% vs.\ static-KB's 39.8\,\%; the guidance instructs the agent to ``keep the first patch minimal,'' and Nemotron interprets this literally. For django-11276 it adds the import without connecting it, for django-14238 it produces a shallow one-liner, and for sympy-11618 it imports \texttt{zip\_longest} without using it.
The root cause is instruction-following interference: Nemotron's different behavioral profile leaves it more susceptible to behavioral steering by prescriptive language, shifting it from executing to deliberating.
Qwen absorbs the same guidance as context while maintaining action-oriented behavior; Nemotron cannot.
The probe-refined condition partially recovers over static-KB (27.0\,\% vs.\ 24.6\,\%), likely because the degraded loop removes some of the most harmful prescriptive language rather than adding useful repo-specific content.

\paragraph{Cross-Model Guidance Transfer Is Catastrophic.}

Qwen-tuned guidance applied to Nemotron does not merely fail to help: it actively collapses agent performance, while leaving per-patch precision intact.
To test whether high-quality guidance produced by a capable model can transfer to a less capable one, we applied Qwen's probe-refined guidance to Nemotron without modification (Table~\ref{tab:transfer-breakdown}).
The result is a collapse: 66 resolved instances (13.2\,\%), compared with 135 (27.0\,\%) for Nemotron's own probe-refined baseline and 142 (28.4\,\%) for no-context.

\begin{table}[t]
\centering
\caption{Agent behavior under Qwen-to-Nemotron cross-model transfer vs.\ Nemotron self-tuned baseline. Qwen's guidance triggers compliance by analysis rather than action in Nemotron, collapsing agent-loop patch production and generating a large fallback cascade. Per-patch precision when the loop completes is unchanged (bottom row).}
\label{tab:transfer-breakdown}
\small
\begin{tabular}{lcc}
\toprule
\textbf{Metric} & \textbf{Qwen guidance} & \textbf{Nemotron baseline} \\
\midrule
Resolved                               & 66 (13.2\,\%)  & 135 (27.0\,\%) \\
Agent-loop patches produced            & 174            & 351            \\
Fallback patches triggered             & 324            & 3              \\
Avg.\ steps / instance                 & 81.0           & 39.2           \\
Instances hitting max steps            & 83             & 11             \\
Eval errors (patch-apply failures)     & 246            & 14             \\
No-command prose events                & 2{,}141        & 1{,}252        \\
Repeated-command stall instances       & 172            & 41             \\
\midrule
Agent-loop precision (resolved/patch)  & 37.9\,\%       & 38.5\,\%       \\
\bottomrule
\end{tabular}
\end{table}

\subsection{Analysis}
The transfer collapse is the most striking of the three findings, so we analyze its mechanism in detail before drawing the cross-cutting lesson.
The mechanism is what we call compliance by analysis: rather than running commands, Nemotron reads Qwen's detailed, prescriptive guidance and writes out an analysis of what it \emph{would} do, producing 71\,\% more turns that issue no command and stalling (repeating itself) four times as often.
This triggers a fallback cascade: when the agent loop produces no patch, the single-shot fallback fires carrying 81 steps of accumulated but unfocused context; the fallback then generates patches that are long but malformed (mean 901 characters versus 280 for the baseline), 243 of 246 evaluation errors are patch-apply failures originating from these fallback patches, and none of the 324 fallback patches resolve the target issue.
Crucially, when the loop does complete and produce a patch, that patch resolves its issue just as often as before: 37.9\,\% under Qwen's guidance vs.\ 38.5\,\% for the Nemotron baseline, a difference well within noise.
The guidance breaks the agent's control loop, producing far fewer evaluable patches, without touching the quality of the patches it does produce.
This is the precision-constancy finding from our main experiment (Section~\ref{sec:coverage}) reappearing from an unexpected direction: across models, across guidance conditions, and under cross-model transfer, the fraction of patches that resolve once they reach evaluation stays stable.
The catastrophe is entirely in coverage.

These results establish that guidance encodes model-specific behavioral calibration, not transferable repository knowledge.
The gains probe-and-refine produces on Qwen reflect guidance tailored to Qwen's specific failure modes; the same text is actively harmful to a model with a different behavioral profile.
This pattern parallels \citet{cloud2025subliminal}'s finding that subliminal transmission of behavioral traits via training data succeeds between models sharing initialization but collapses across families. Behavioral-calibration artifacts, whether produced by weight updates, training data, or prompt-level tuning, appear to encode model-family-specific signals that do not transfer cleanly.
Whether guidance explicitly designed for cross-model use, or tuned to avoid model-specific behavioral cues, could support positive transfer is an open question; this experiment tests only incidental transfer, and a targeted multi-model tuning regime might not trigger the same failure mode.

\section{Discussion}
\label{sec:discussion}

\paragraph{Instruction quality as a first-order variable.}
The central claim of this paper is that the quality of the instructions given to a coding agent is a first-order determinant of its reliability, comparable in magnitude to choices about model capability or step budget within a fixed scaffold.
This reframes a question that has received surprisingly little systematic attention: while models, context windows, and scaffolds have all been subject to careful ablation, the content of the agent's own instructions has rarely been varied or optimized.
Guidance does not improve the quality of the changes the agent makes (precision is constant at $\sim$59\,\%), but, as the budget experiment shows (Section~\ref{sec:budget}), it is what makes additional steps productive at all.
The 42\,\% of instances resistant to all three conditions at the any-trial threshold are unlikely to be reached by guidance improvements alone; these appear to require stronger reasoning or longer context rather than better localization.

\paragraph{A speculative interpretation: prompt-level cluster activation.}
A more speculative reading of these results connects them to recent work on narrow fine-tuning.
\citet{betley2025emergent} report that narrow weight updates can produce broad behavioral change, and \citet{betley2025weird} provide mechanistic evidence, via a sparse-autoencoder analysis in one of their experiments, that this happens by strengthening broad persona features already present in the model rather than narrow task-specific ones.
Probe-and-refine's empirical signature is similar. Ten synthetic probes per iteration shape an agent's behavior on hundreds of unrelated SWE-bench instances, and the added guidance is dominated by procedural workflows the model could in principle execute when prompted (Table~\ref{tab:content-analysis}).
One reading is that the procedure does not teach the agent new capabilities but locates and activates dormant ones from outside the weights.
We make no mechanistic claim for the prompt-space case. The present results are consistent with this interpretation but do not test it directly, and a representation-level or probing study analogous to \citet{betley2025weird}'s would be required to distinguish it from alternatives.

\paragraph{Reconciling prior findings.}
\citet{lulla2026agentsmd} evaluate focused pull requests where budgets are likely generous relative to task complexity; \citet{gloaguen2026evaluating} evaluate complex benchmark tasks under default settings.
Our budget experiment shows that the same guidance can appear beneficial, neutral, or harmful depending on available steps.
Additionally, \citet{gloaguen2026evaluating}'s context files are generated in a single LLM pass, while probe-and-refine guidance is iteratively refined through failure feedback, consistent with \citet{zhu2026swecontextbench}'s finding that curated experience helps while unfiltered experience does not.
Critically, the Nemotron self-tuned experiment provides a within-study replication of \citet{gloaguen2026evaluating}'s finding: the same \texttt{static\_kb} guidance that adds 2.8\,pp to Qwen's resolve rate reduces Nemotron's by 3.8\,pp.
The disagreement between prior studies may reflect differences in the models evaluated as much as differences in how guidance was produced or what tasks were attempted.
To our knowledge, this is the first systematic variation of step budget across guidance conditions in the SWE-bench literature; existing studies report a single fixed budget without examining how it interacts with the guidance provided.

\paragraph{Model capacity as a moderating variable.}
Practitioners should treat guidance complexity as a three-way match: workflow depth, step budget, and model capacity must all be commensurate for probe-and-refine to help rather than harm.
A capacity-constrained model cannot sustain the feedback loop (Section~\ref{sec:nemotron}), and prescriptive guidance calibrated for a more capable model triggers mode-switching rather than execution.

\paragraph{Guidance is model-specific.}
The cross-model transfer experiment (Section~\ref{sec:nemotron}) demonstrates that guidance encodes model-specific behavioral calibration.
Practitioners should tune guidance with the model that will consume it.

\paragraph{Contamination robustness.}
\label{sec:contamination}
The 31 unique consistent solves from probe-and-refine are distributed across 9 of 12 repositories (Figure~\ref{fig:localization-repos}) and mirror the overall difficulty distribution, suggesting the improvement comes from guidance content rather than memorization.
We cannot rule out that guidance triggers recall of memorized solutions the unguided prompt fails to elicit, but the unique solves show no skew toward older or more-starred issues.

\paragraph{Absolute performance.}
Resolve rates (22--35\,\%) reflect a constrained pipeline: a 35B mixture-of-experts model with 16k effective context and aggressive output truncation.
Frontier systems achieve $>$55\,\% on SWE-bench Verified.
The contribution is a controlled experiment showing that iteratively refined guidance significantly outperforms alternatives within the same pipeline.

\section{Limitations}
\label{sec:limitations}

\paragraph{Variance characterization.}
Each condition was run in four independent trials at 200 steps, allowing us to characterize but not eliminate run-to-run variance.
Additional trials would tighten confidence intervals further but are unlikely to change conclusions: the hierarchical model already gives $p < 0.001$ on the key contrasts at $n{=}4$ trials.
The lower-budget runs (25, 50, 100 steps) in Section~\ref{sec:budget} are single-trial measurements and lack the same uncertainty quantification.

\paragraph{Guidance length confound.}
The refined guidance is 63\,\% longer on average than the static-KB guidance (Table~\ref{tab:char-stats}), so we cannot fully disentangle content quality from prompt length.
The content analysis (Table~\ref{tab:content-analysis}) shows that the added material is predominantly repo-specific knowledge (procedural guardrails, structural references, and quality gates) rather than generic padding.
Furthermore, \citet{gloaguen2026evaluating} demonstrate that longer LLM-generated context files can actively \emph{reduce} performance, which is the opposite of what a pure length effect would predict.
The localization analysis (Section~\ref{sec:localization}) provides additional support: the instances probe-refined guidance uniquely resolves are concentrated in repositories where naive symbol-based localization fails, a pattern that generic-text padding would not produce.
Two ablations would isolate this confound completely (padding \texttt{static\_kb} to match refined-guidance length, and truncating refined guidance to \texttt{static\_kb} length); we did not run these due to compute constraints, and the possibility that the model simply benefits from additional prompt tokens remains open.

\paragraph{Single model with demonstrated capability.}
The probe-and-refine procedure has been demonstrated on one model, Qwen3.5-35B-A3B.
A cross-model experiment with Nemotron suggests the procedure has model-fit requirements: the model must generate diagnostic output rich enough for the judge step to act on, though the loop prompts were designed around Qwen's output format and a Nemotron-adapted design remains untested.
Nemotron's loop degrades and even static guidance hurts rather than helps it on this scaffold.
The open question is whether the procedure generalizes to a second model that can run it: a larger dense model with strong tool-use behavior, for instance.
Scaffolds with context summarization or tree-search may interact with guidance differently regardless of model capacity.

\paragraph{Probe sensitivity.}
We generate 10 probes per iteration at temperature 0.9 with deduplication and provide a qualitative analysis of what the procedure discovers (Section~\ref{sec:par-method}), but do not conduct a systematic sensitivity analysis of how the final guidance changes under different probe sets or random seeds.
Whether the procedure converges to similar guidance from different random seeds is an open question; the early-stop behavior (3--5 iterations) suggests some runs may be more productive than others.

\paragraph{Single benchmark with Django concentration.}
Findings are specific to SWE-bench Verified (500 instances), with Django accounting for 46\,\%.
Table~\ref{tab:django-split} shows the effect holds in both subsets, but the non-Django subset contains 11 repositories with as few as 1--2 instances each, making per-repo significance testing infeasible and limiting claims about cross-repository generality.
The effect could be primarily a Django effect with modest support from other repositories; evaluation on benchmarks with more balanced repository distributions is needed.

\section{Conclusion}
\label{sec:conclusion}

We introduced probe-and-refine tuning, a lightweight procedure that iteratively refines a statically generated repository knowledge base into specialized operational guidance through $\sim$22 single-shot LLM calls per iteration: generate a batch of probes, attempt and judge each probe against expected behaviors, and aggregate diagnostics into mechanically applied guidance edits.
Across four independent trials on SWE-bench Verified, the resulting guidance achieves a mean resolve rate of 33.0\,\% (mixed-effects logistic regression $p < 0.001$ vs.\ both baseline and static-KB) and produces 31 instances solved consistently ($\geq 3$ of 4 trials) that no other condition resolves.
The mechanism is evaluation coverage: the guidance gives the agent a structured workflow that converts exploration into well-formed, evaluable patches more reliably, without improving the correctness of any individual evaluated patch.
A qualitative analysis suggests the procedure functions as a localization aid, helping the agent reach the correct file for small mechanical fixes, rather than enabling fundamentally harder repairs.

The procedure is simple enough to be surprising.
That a procedure composed entirely of single-shot LLM calls ($\sim$22 per iteration), with no tool use and no multi-step reasoning, can encode repo-specific operational knowledge sufficient to significantly outperform both an unguided agent and a static knowledge base suggests that the instructions given to coding agents matter at least as much as the agents' own reasoning capabilities.

A cross-model analysis with Nemotron reveals a model-fit dependency: the procedure appears to require a model that can produce rich diagnostic output and judge its own patch attempts with enough specificity to drive the feedback loop.
Below this threshold the loop degrades, guidance interferes rather than helps, and the precision-constancy finding (that per-patch quality is unaffected by guidance) still holds, now replicated under cross-model transfer.

For practitioners: invest in iteratively refined guidance for repositories where your agent will handle many issues, ensure the step budget accommodates the prescribed workflow, and tune guidance with the model that will consume it, as guidance calibrated for a different model's behavioral profile may actively harm performance.

\paragraph{Author Contributions.}
A.S. designed and ran the experiments, performed the analysis, and wrote the manuscript. J.A. supervised the project and provided feedback on the methodology and manuscript.


\end{document}